\documentclass[a4paper,11pt,article,oneside]{memoir}
\usepackage{ecis2024}
\usepackage{hyperref}

\maintitle{Scaffolding Creativity: How Divergent and Convergent LLM Personas Shape Human Machine Creative Problem-Solving} 
\shorttitle{Divergent and Convergent LLM Personas} 
\category{}

\authors{
	Alon Rosenbaum, Ben-Gurion University of the Negev, Israel, alonr91@gmail.com
	
	Yigal David, Shenkar -  College of Engineering. Design. Art, yigal.david@shenkar.ac.il
	
	Eran Kaufman, Shenkar -  College of Engineering. Design. Art, erankfmn@gmail.com
	
	Gilad Ravid, Ben-Gurion University of the Negev, Israel, rgilad@bgu.ac.il
	
	Amit Ronen, Shenkar -  College of Engineering. Design. Art, amitronen222@gmail.com
	
	Assaf Krebs, Shenkar -  College of Engineering. Design. Art, assaf.krebs@shenkar.ac.il
}

\shortauthors{Rosenbaum et al.} 

\begin{document}

	\begin{abstract}\noindent
		Large language models (LLMs) are increasingly shaping creative work and problem-solving; however, prior research suggests that they may diminish unassisted creativity. To address this tension, a coach-like LLM environment was developed that embodies divergent and convergent thinking personas as two complementary processes. Effectiveness and user behavior were assessed through a controlled experiment in which participants interacted with either persona, while a control group engaged with a standard LLM providing direct answers.
		
		Notably, users' perceptions of which persona best supported their creativity often diverged from objective performance measures. Trait-based analyses revealed that individual differences predict when people utilize divergent versus convergent personas, suggesting opportunities for adaptive sequencing. Furthermore, interaction patterns reflected the design thinking model, demonstrating how persona-guided support shapes creative problem-solving.
		
		Our findings provide design principles for creativity support systems that strike a balance between exploration and convergence through persona-based guidance and personalization. These insights advance human-AI collaboration tools that scaffold rather than overshadow human creativity.
	\end{abstract}
	
	
	\chapter{Introduction}\label{introduction}
	
	Introduction
	Large language models (LLMs) are rapidly being integrated into information systems supporting ideation, problem framing, and solution development across organizational contexts. Despite their broad adoption, the impact of LLMs on human creativity remains unsettled. While structured, coach-like AI assistance can aid users in exploring and refining ideas, generic answer provision risks homogenizing outputs, anchoring users to initial suggestions, and narrowing the search space \citep{frich2021digital, shaer2024ai, suh2024luminate}. Consequently, the critical design challenge is not whether to employ LLMs in creative knowledge work but how to structure interactions so that these systems scaffold exploration and evaluation without diminishing human judgment and agency \citep{bansal2019beyond, jakesch2023co}.
	
	Creative methodologies such as Design Thinking emphasize an iterative alternation between divergence, expanding options and prompting novel ideas, and convergence, evaluating and selecting among possibilities. The Double Diamond framework formalizes this cadence and cautions against premature convergence that suppresses originality \citep{design2019framework, brown2008design, meinel2011design, cropley2006praise}. However, most LLM interfaces conflate these modes within a single chat environment, blurring task phases essential for effective creativity. Human-computer interaction (HCI) research suggests that making these mode shifts explicit can improve collaborative processes, yet controlled evidence testing whether exposing divergent and convergent modes in LLM interfaces alters creative processes and outcomes remains scarce \citep{frich2021digital}.
	
	Three key gaps persist in IS and HCI literature. First, few studies implement and compare interfaces that instantiate both divergent and convergent personas, allow users to freely switch between modes, and jointly analyze creative process dynamics, including curiosity and switching behavior, and product-level originality and diversity. Second, users' perceptions of AI-enabled creativity support often diverge from objective success metrics, with this misalignment potentially affecting trust calibration and adoption; however, many works do not simultaneously capture subjective evaluations alongside behavioral traces \citep{bansal2019beyond, jakesch2023co}. Third, the moderating role of individual differences, particularly personality traits known to influence creative styles (e.g., openness linked to divergence, conscientiousness to persistence), is underexplored in the context of persona-guided LLM support for creative problem-solving \citep{mccrae1987creativity, cropley2006praise}.

	The remainder of the paper reviews related work, describes the system and experimental design, reports results, and discusses implications for the design and governance of creativity support systems.
		
	\hypertarget{background-related-work}{
		\chapter{Background \& Related Work}\label{background-related-work}}
	
	\section{Creative Process and Mode Switching}
	Creativity research consistently positions ideation as cycling between divergent exploration, generating many varied possibilities, and convergent evaluation, refining and selecting among alternatives \citep{Guilford1967, cropley2006praise, Simonton2015, baer2014creativity}. Classic creativity models advise decoupling these phases to avoid premature narrowing of ideas and to promote richness of outcome \citep{Wallas1926, frich2021digital}. Brainstorming exemplifies this by privileging idea fluency prior to judgment \citep{osborn1953applied}.
	
	At the interpersonal level, collaboration unfolds in a dialogic space where cognitive diversity and intersubjectivity enable shared meaning-making \citep{Wegerif2013}. Mode-switching involves opening a shared possibility space, broadening perspectives through additional voices, and deepening reflection and critique \citep{Wegerif2013}. Technology can scaffold these shifts by introducing alternative viewpoints or information sources \citep{HirvonenPalmgrenNeuvonen2019}. While nominal groups may suffer from production blocking, committed teams benefit from cognitive stimulation and exposure to diverse ideas \citep{nijstad2003production, dugosh2000cognitive}.
	
	The transition from divergence to convergence is encapsulated as interthinking, collective reasoning that connects and refines ideas through dialogue rather than mere idea listing \citep{LittletonMercer2013}. Open-ended creative collaboration involves intercreating, wherein participants co-construct problem framings and constraints to explore novel possibilities \citep{palmgren2017intercreating}. Empirical work shows that engagement in dialogue integration better predicts novel final outputs than idea quantity alone \citep{coursey2019linking}.
	
	Practical facilitation of creative dialogue leverages role framing, rotation, visible norms, and balanced turn-taking, alongside prompts for alternative views to maintain productive flow without overshadowing participant voices \citep{PaulusBrown2003, Wegerif2013, Dillenbourg2013}. Models such as the Double Diamond and C–K theory operationalize these dynamics in formal innovation workflows \citep{tschimmel2012design, hatchuel2003new}, emphasizing iterative divergence-convergence cycles.
	
	\section{Personality Traits and Creativity}
	Creativity arises within a complex system where individual personality traits interact with social constraints and task demands \citep{hennessey2010-creativity, ford1996-theory-individual}. The dual-pathway model proposes creativity stems from flexible broad search or persistent deep search, modulated by affect and context \citep{nijstad2010-dual-pathway}. Therefore, trait influences on creativity may be phase and context dependent.
	
	The Big Five personality framework, openness, conscientiousness, agreeableness, extraversion, and neuroticism, is widely used to capture trait effects on creativity \citep{grajzel2023-big-five}. Openness consistently correlates positively with creative potential and divergent thinking \citep{kaufman2016openness}. The link for conscientiousness is nuanced, with facets producing opposing relations \citep{reiter-palmon2009-conscientiousness}. Extraversion shows moderate positive associations with divergent creativity, while agreeableness and neuroticism effects are context-dependent \citep{feist1998-meta-analysis}. In social creativity, leader traits such as agreeableness may negatively affect critical feedback efficacy \citep{grajzel2023-big-five}.
	
	Creativity evaluation standards combine originality and effectiveness \citep{runco2012standard}. Psychometric tools measure fluency, flexibility, originality, and elaboration \citep{torrance1966-torrance-tests}. Expert judgment techniques provide domain-valid assessments but are labor intensive \citep{baer2009assessing, amabile1983social}. Within HCI and IS, objective fluency metrics (e.g., unique idea counts) complement subjective reports to triangulate creative outcomes \citep{hewett2005-creativity-support, remy2020evaluating}.
	
	\section{Large Language Models and Personas}
	LLMs based on Transformer architectures \citep{vaswani2017attention} underpin state-of-the-art natural language generation but pose challenges including output controllability, bias risks, and user homogenization \citep{anderson2024homogenization, zamfirescu-Pereira2023-johnny}. Persona construction, through prompts, system roles, memory, and tool integration, enables behavioral steering and role specialization in conversational agents \citep{liu2023pre, shaer2024ai}. Studies show that explicit role assignments improve multi-turn interaction coherence and role fidelity \citep{Qin2024characterMeet, schwabe2025ai}.
	
	Personas function as calibrated partners or coaches rather than mere answer providers, emphasizing augmentation of human judgment over substitution \citep{hofman2023steroids, collins2024building}. Risk mitigation includes transparency, role disclosure, and interface cues to prevent over-trust and automation bias \citep{vered2023effects}.
	
	\section{Human-LLM Co-Creative Ideation}
	Persona-guided LLM systems facilitate dialogic expansion, broadening exploration and supporting collective evaluation \citep{liu2025personaflow, Wegerif2013}. Research on multi-persona or multi-agent configurations reveals enhanced fluency, novelty, and engagement but mixed effects on relevance and dependence \citep{kumar2025human, specker2025extra, shaer2024ai, wan2025using}. Persona diversity serves as a hedge against group homogenization biases when properly integrated.
	
	Although many experiments demonstrate promising outcomes, limitations regarding sample sizes, task generalizability, and external validity remain \citep{liu2025personaflow, shaer2024ai}. This study addresses these gaps by experimentally manipulating persona-guided divergent and convergent modes, modeling individual traits, and employing computational-semantic originality metrics within an IS context.
	
	\hypertarget{research questions}{
		\chapter{Research Questions}\label{researchquestions}}
	
	This study investigates how persona-guided large language model (LLM) interfaces influence creative collaboration, user perception, and behavioral engagement within a socio-technical system. Drawing on theories of divergent and convergent thinking as complementary modes of creativity, the research examines how distinct AI personas affect user experience, creativity outcomes, and individual differences in interaction dynamics. The work is guided by three overarching research questions: 
	\\
	\\    
	\textbf{RQ1:} How does persona-guided interaction, compared to a standard LLM interface, affect users’ perceived creativity enhancement and ownership of ideas during a co-creative problem-solving task?
	\\
	\\
	\textbf{RQ2:} To what extent do individual differences in personality traits (e.g., openness, agreeableness, conscientiousness) shape user attitudes, engagement patterns, and preferences toward divergent versus convergent AI personas?
	\\
	\\  
	\textbf{RQ3:} How do persona-guided LLM interactions influence the originality and diversity of user-generated ideas, as measured through computational semantic analysis of conversation-based ideation?
	\\
	\\
	Collectively, these questions aim to bridge human–computer interaction, creativity research, and information systems design by assessing whether persona differentiation in generative AI systems can systematically modulate creative outcomes and user experience.

	\hypertarget{methodology}{
		\chapter{Methodology}\label{methodology}}
	
	A total of 105 undergraduate students from a design and engineering institution participated in a randomized experiment. Participants were compensated \$15 each, with a \$150 voucher awarded for the highest-rated submission. Random assignment allocated 73 participants to the treatment group and 32 to the control group.
	
\begin{figure}
		\centering
		\includegraphics[width=0.6\linewidth]{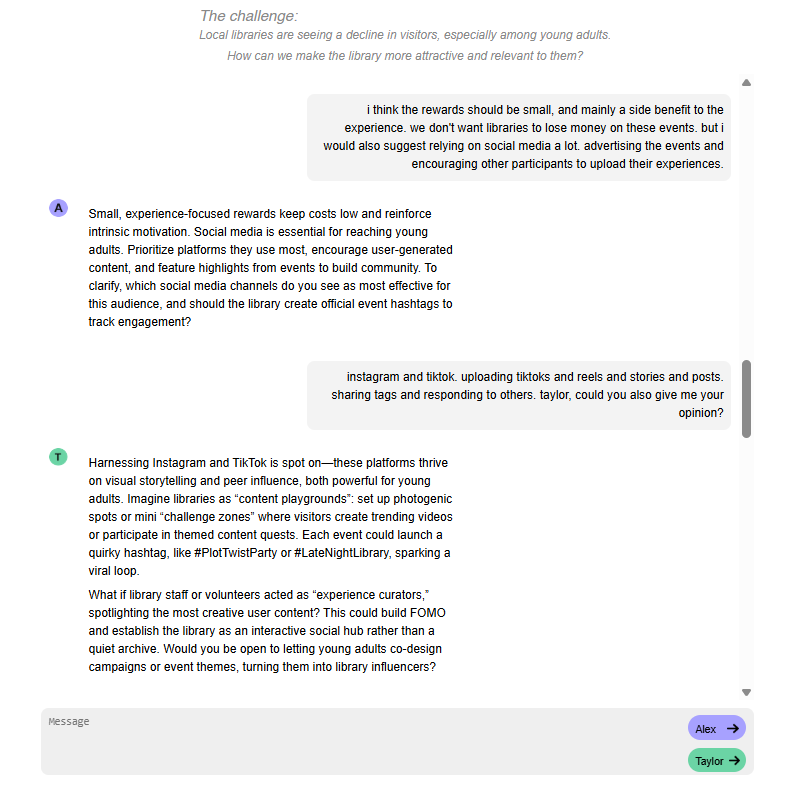}
		\caption{Persona-Guided Chat Interface used in the experiment.}
		\label{fig:ui}
	\end{figure}

	Before task initiation, participants completed a standardized five-minute orientation that introduced the experimental interface (Figure \ref{fig:ui}). The interface replicated a typical large language model (LLM) chat environment, presenting two action buttons labeled “Taylor” (purple) and “Alex” (green). Participants were informed that selecting a button directed input to the respective agent, but the underlying distinctions between agents were intentionally undisclosed to maintain blinding.
	
	Following orientation, all participants engaged in the same creative problem-solving task: “How can we make libraries more attractive to young adults?” Each participant had 20 minutes to complete the task after submitting digital consent and accessing the system interface.
	
	In the control condition, both buttons routed inputs to the same baseline ChatGPT model, yielding unmodified responses regardless of persona selection. In the treatment condition, differentiated prompt-engineering strategies were implemented. The “Taylor” persona was optimized for divergent thinking using envelope prompts and a higher temperature parameter (0.8) to promote variability and idea fluency. The “Alex” persona was configured to support convergent thinking, employing structured prompts and a lower temperature (0.3) to produce focused, analytical responses.
	
	To preserve continuity and minimize persona drift, the system employed stateful context management. Each model query packaged a static system prompt, a dynamically generated JSON summary of the conversation state, and the transcript of recent exchanges. This payload construction follows best practices for persistent context in multi-turn, agent-based LLM environments.
	
	The interface incorporated several bias-mitigation mechanisms illustrated in Figure \ref{fig:ui}: gender-neutral persona names, visually balanced color assignments, and randomized vertical placement of send buttons to counter positional effects. System telemetry automatically logged conversation content, message timestamps, persona switching patterns, and total interaction length for subsequent analysis.
	
	Recognizing the relative performance advantages of LLMs in English, non-native participants were allowed to use an online Translator before submission.
	
	Upon task completion, participants filled out a post-session questionnaire including the Big Five Inventory-2-XS (BFI-2-XS) \citep{soto2017short}), demographic information, and subjective reflections on the dialogue experience. All questionnaire items were rated on a five-point Likert scale, as summarized in Table \ref{table:quest}. The protocol received ethics approval from the departmental research committee of the first author’s institution.
	
	\begin{table}
		\begin{tabular}{||l||}
			\hline
			1. Alex helped me arrive at a creative solution \\
			2. Taylor helped me arrive at a creative solution \\
			3. My creativity increased compared to my usual level after chatting with Alex \\
			4. My creativity increased compared to my usual level after chatting with Taylor\\
			5. The solution originated from me \\
			6. Compared to regular ChatGPT, the interface helped me reach a creative solution \\
			7. Perceived proficiency with generative AI tools (e.g., ChatGPT, Gemini, Claude) \\
			8. Which persona most enhanced your creativity? (forced-choice: 1=Taylor ... 4=Alex)\\
			\hline
		\end{tabular}
		\caption{Perceived Attitude Toward Dialogue Questionnaire.}
		\label{table:quest}
	\end{table}

	\hypertarget{results}{
		\chapter{Results}\label{results}}
	Out of 105 participants, four were excluded due to insufficient system engagement: minimal interaction (n=1), inadequate session duration (n=2), session timeout violations (n=1), and content authenticity issues (n=1). The final dataset comprised 101 conversation logs: 69 from the persona condition and 32 from the control condition. An additional three participants per group were excluded from survey analysis due to incomplete questionnaire responses.
	
	Participants averaged 27.65 years of age with no significant between-condition differences. Academic distribution was balanced across groups: 50\% design students, 30\% engineering students, 4\% business majors, and 16\% from science, humanities, or other disciplines. Welch t-tests confirmed no significant baseline differences in Big Five personality traits between conditions.
	
	\begin{figure}
		\centering
		\includegraphics[width=0.7\linewidth]{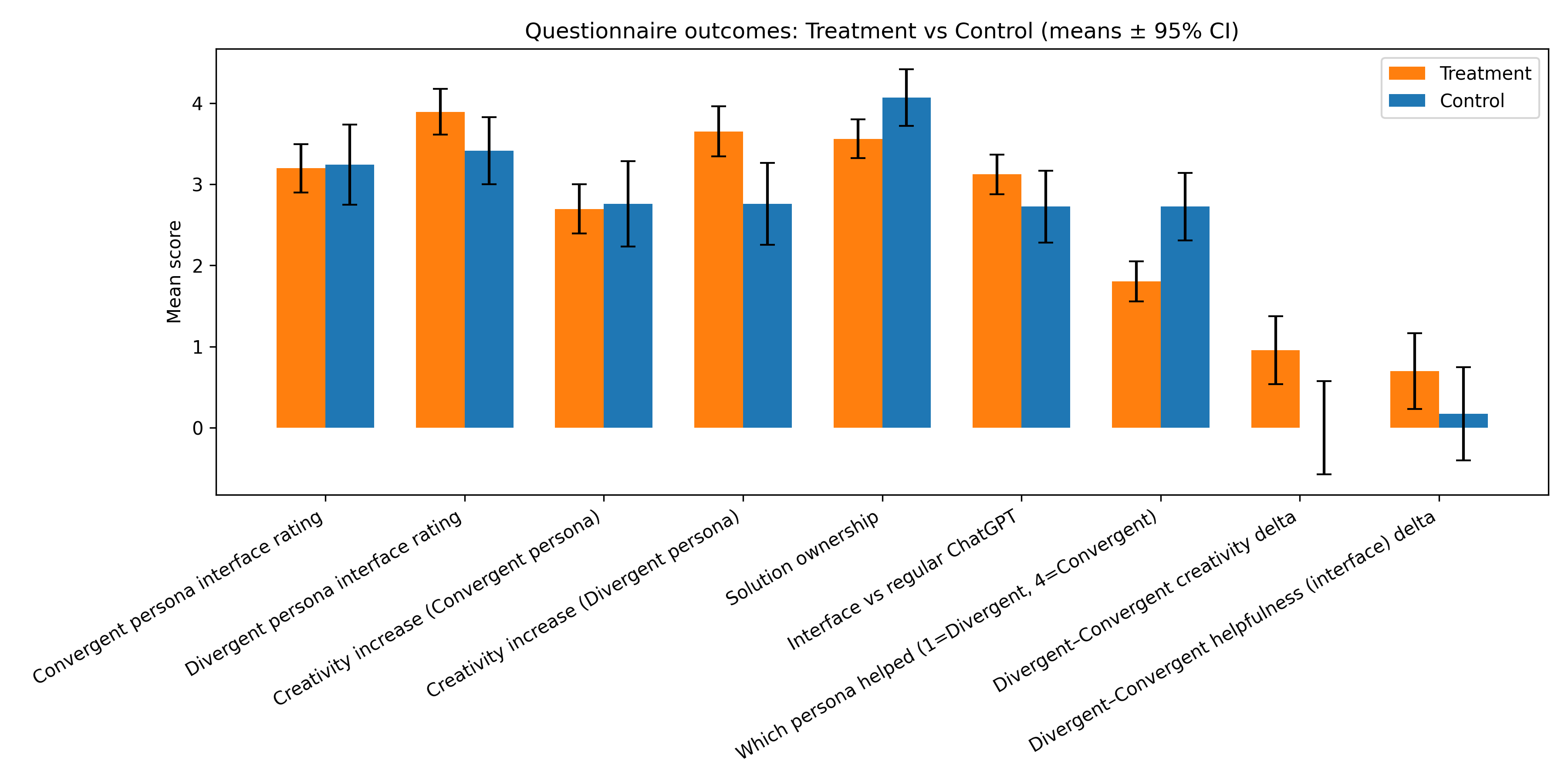}
		\caption{Post-session questionnaire outcomes by experimental condition.}
		\label{fig:questExp}
	\end{figure}
	
	Post-session evaluations revealed significant between-condition differences in system perception and persona utility (Figure \ref{fig:questExp}). Statistical analyses employed Welch unequal-variances t-tests with Hedges' g effect size reporting.
	
\begin{figure}
		\centering
		\includegraphics[width=0.7\linewidth]{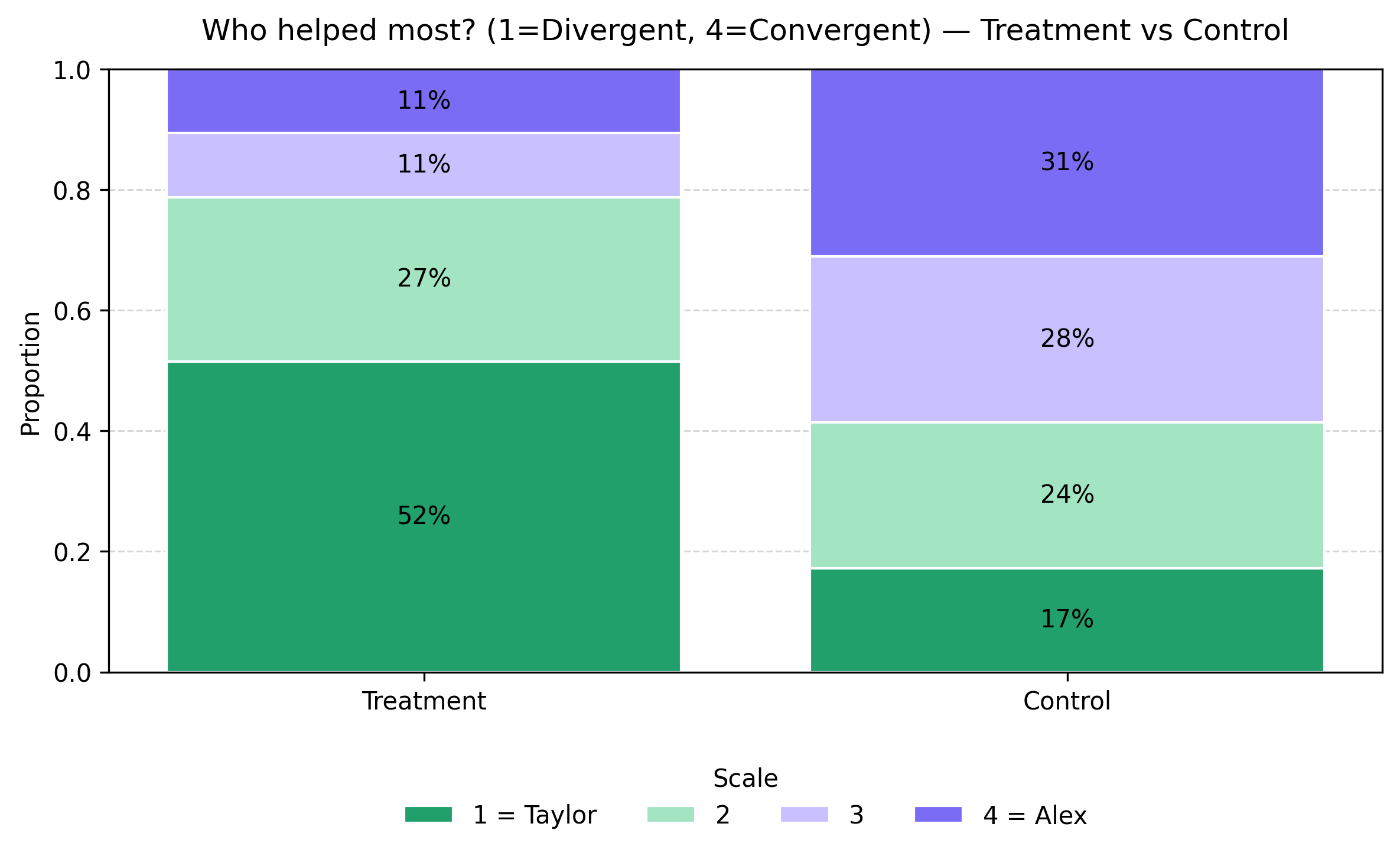}
		\caption{Distribution of persona preference ratings (1-4 scale).}
		\label{fig:whoHelped}
	\end{figure}

	On the forced-choice item "Which persona most enhanced your creativity?" (scale: 1=divergent, 4=convergent)(item 8 table \ref{table:quest}), treatment participants showed stronger preference for the divergent persona (Treatment M=1.80, SD=1.01; Control M=2.72, SD=1.10; $\Delta$=-0.92, g=-0.88, p=.0003). Response distributions demonstrated clear polarization: treatment responses concentrated on the divergent side (52\% rated 1, 27\% rated 2; total Taylor-preference=78.8\%), while control responses favored the convergent side (28\% rated 3, 31\% rated 4; total Alex-preference=58.6\%) as shown in Figure \ref{fig:whoHelped}. One-sample tests against the scale midpoint (2.0) revealed significant convergent preference in the control condition (t(28)=3.55, p=.0013) but no significant divergent preference in treatment (t(65)=-1.58, p=.118). This pattern indicates a between-condition shift toward crediting the divergent persona rather than definitive within-treatment preference below neutrality.
	
	For creativity enhancement ratings (items 1 \& 2 table \ref{table:quest}), the divergent persona (Taylor) received significantly higher scores in the treatment condition (Treatment M=3.65, SD=1.26; Control M=2.76, SD=1.33; $\Delta$=+0.89, g=0.69, p=.003). The convergent persona (Alex) showed no between-condition differences (Treatment M=2.70, SD=1.23; Control M=2.76, SD=1.38; p=.837).
	
	Within-participant selectivity analysis using Taylor-minus-Alex difference scores revealed stronger divergent preference in treatment participants. On creativity enhancement, the delta was significantly larger in treatment (M=0.95, SD=1.70) than control (M=0.00, SD=1.51; g=0.58, p=.008). For solution helpfulness, treatment participants also showed numerically higher divergent preference (M=0.70, SD=1.90) compared to control (M=0.17, SD=1.51), though this difference was not statistically reliable (g=0.29, p=.156).
	
	The persona-guided interaction resulted in reduced perceived solution ownership . Treatment participants were less likely to agree that "the solution came from me" (item 5 table \ref{table:quest}) (Treatment M=3.56, SD=0.98; Control M=4.07, SD=0.92; $\Delta$=-0.51, g=-0.52, p=.018), suggesting that persona differentiation redistributes authorship attribution from individual to collaborative processes.

	\begin{figure}
		\centering
		\includegraphics[width=0.7\linewidth]{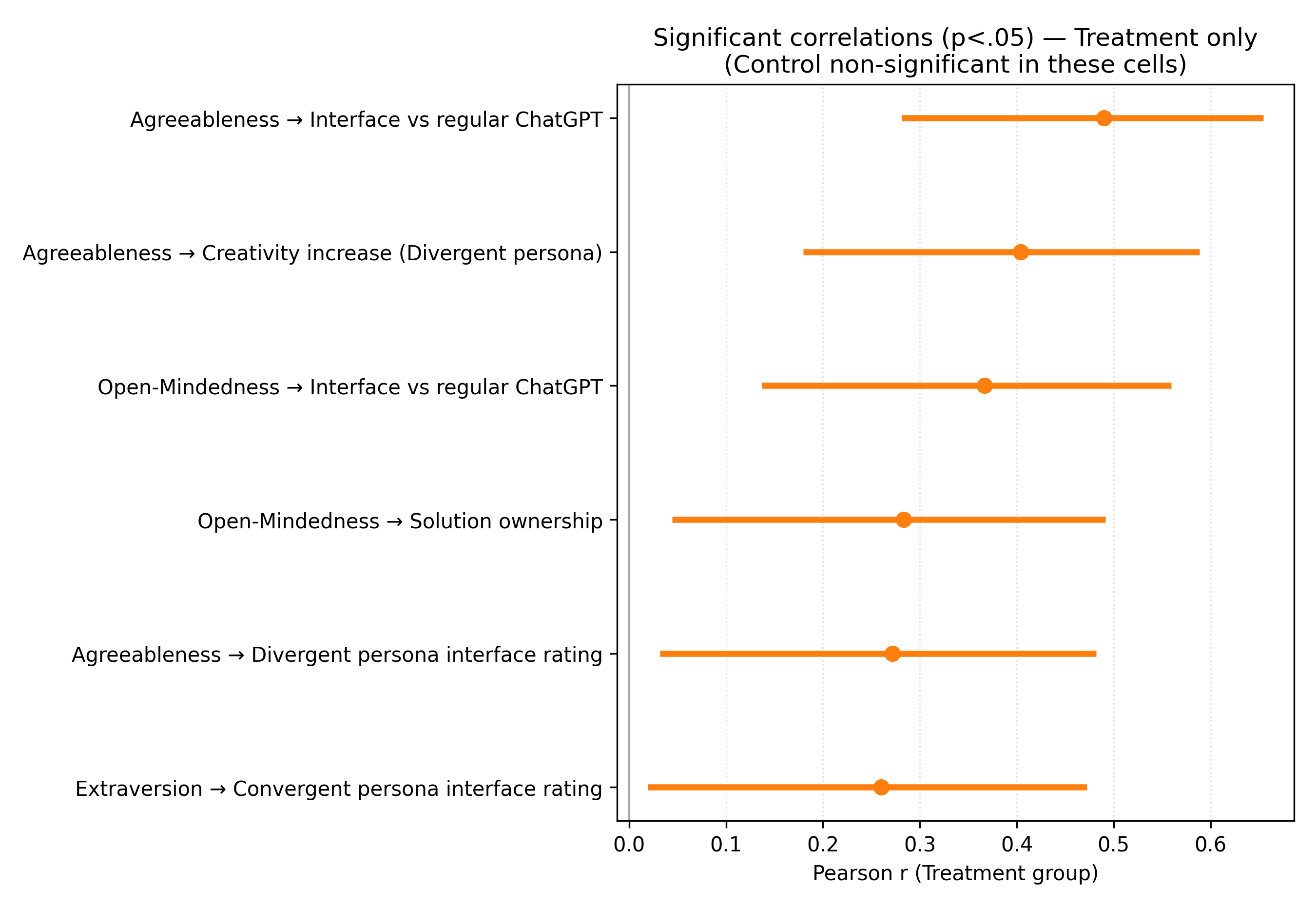}
		\caption{Significant personality trait-outcome correlations in the treatment condition. Error bars represent 95\% confidence intervals around Pearson's r.}
		\label{fig:big5}
	\end{figure}
	
	Big Five personality traits significantly predicted system attitudes within the treatment condition, while control condition correlations remained non-significant (all p>.14) as shown in Figure \ref{fig:big5}. 
	
	Agreeableness (M=4.06, SD=0.62) emerged as the strongest predictor of positive persona evaluation, correlating with higher divergent persona ratings (r=0.27, p=.027), greater perceived creativity enhancement from the divergent persona (r=0.40, p<.001), and superior performance relative to standard ChatGPT (r=0.49, p<.001). Extroversion (M=3.32, SD=0.82) predicted convergent persona preference (r=0.26, p=.035). Openness (M=3.84, SD=0.77) correlated positively with system superiority over ChatGPT (r=0.37, p=.002) and solution ownership retention (r=0.28, p=.021).

	\section{User Interaction and Engagement}
	
	Behavioral analysis of conversation dynamics revealed systematic personality-driven usage patterns. Participants higher in conscientiousness directed more messages toward the convergent persona (r=0.297, p=.016) and sustained longer convergent interaction sequences (r=0.363, p=.003). These associations were absent in the control condition (both p>.18).
	
	Quartile-based analysis of conscientiousness scores demonstrated clear ending-persona preferences. Low-conscientiousness users predominantly concluded sessions with the divergent persona (14/16 vs. 2/16 for convergent), whereas high-conscientiousness users more frequently ended with convergent interactions (10/16 vs. 6/16 for divergent; $\chi^2$(1)=6.533, p=.011). Neuroticism showed similar patterns (high-neuroticism users: 13/16 ended with convergent; low-neuroticism: 6/16 divergent; $\chi^2$(1)=4.664, p=.031). Control condition participants showed no trait-based ending preferences (all p $\geq$ .266).
	
	\section{Information-Seeking Behavior}
	Question-asking frequency served as a behavioral indicator of interaction and engagement. To ensure that the question marks appearing in users’ prompts are indeed aimed at enhancing knowledge and deepening idea development within a contextual dialogue, we conducted a qualitative analysis of the content of user interactions. A very clear picture emerges: the vast majority of question marks (more than 80 in the first quarter, and almost 100
	
	“Can you mention influencers associated with culture, literature and technology in the central region?”
	“What do you think of him as the 17-year-old we described earlier?”
	“Instead of visitor graffiti, I would create… spaces where everyone could find themselves, just like [the central library in the city] — what do you think, Taylor?”
	“What kind of furniture is suitable for all ages for the first idea of study and work?”
	“Everyone experiences the outside differently, and we want to focus the feeling of the library as a pleasant emotional space. What do you think about it?”
	“I’d love your opinion — should the change in libraries ... be gradual or fundamental?”
	“Alex, what do you think about what Taylor and I did?”
	“Can you also add the resources needed for this?”
	“What is your evidence? You provide no clear data to show youth prefer your choice.”
	“These seem pretty costly. Which one of these has the highest ROI?”
	
	The quantitative analysis of question marks focused on quarters 2-4 of conversations to exclude initial familiarization phases. Two metrics captured questioning behavior: mean question marks per user message and percentage of user turns containing questions.
	
	A clear picture emerges from the data analysis, where there is a significant difference in the level of engagement and interaction between the treatment group and the control group.
	
	The divergent persona significantly increased questioning on both measures compared to control baselines (mean question marks: t=3.09, p=.003, $n_treatment$=63, $n_control$=51; percentage with questions: t=2.70, p=.008). The convergent persona also elevated question frequency (mean question marks: t=2.15, p=.034,$n_treatment$=59, $n_control$=51), though the percentage metric showed only a non-significant trend (t=1.56, p=.122).
	
	\begin{figure}
		\centering
		\includegraphics[width=0.7\linewidth]{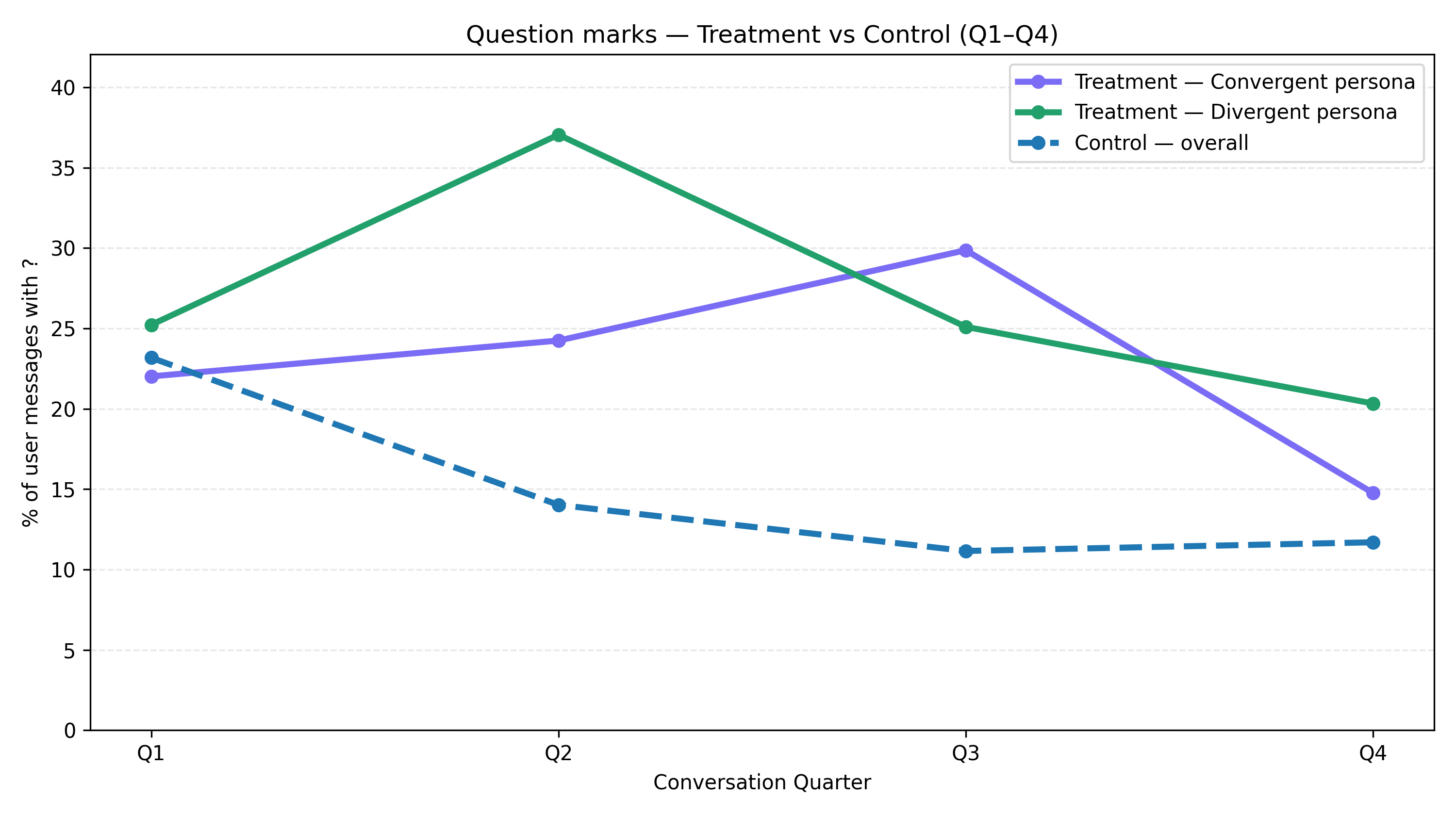}
		\caption{Question mark frequency in mid-to-late conversation phases.}
		\label{fig:qmark}
	\end{figure}
	
	\section{Computational Creativity Assessment}
	The assessment of creativity through its four primary components—fluency, originality, flexibility, and elaboration, has been extensively studied and operationalized in research. These components, originally conceptualized by Guilford \citep{guilford1950creativity} and later refined by Torrance, form the foundation of most contemporary creativity measurement instruments, particularly the Torrance Tests of Creative Thinking (TTCT)\citep{torrance1966torrance}.
	
	Fluency represents the quantitative dimension of creative output and is measured by calculating the total number of relevant responses an individual produces within a given time frame. In divergent thinking tasks, such as generating alternative uses for common objects, fluency scores reflect the sheer volume of appropriate ideas generated. This measure serves as an indicator of ideational productivity and the ability to access and retrieve multiple solutions from one's cognitive repertoire. Research has shown that individuals with higher fluency scores demonstrate greater facility in generating ideas across various domains, though quantity does not necessarily correlate with quality.
	
	Originality constitutes perhaps the most distinctive aspect of creative assessment, focusing on the statistical infrequency or uniqueness of responses. This component is typically evaluated through comparison with normative databases, where responses occurring in less than five percent or one percent of the population receive progressively higher scores. Alternative scoring methods employ trained raters who evaluate the novelty and unusualness of responses based on established criteria. The measurement of originality attempts to capture the essence of creative thinking—the ability to produce ideas that deviate from conventional or common responses while maintaining relevance to the task parameters.
	
	Flexibility measures the breadth of cognitive categories employed in creative responses, indicating an individual's capacity for cognitive shifting and conceptual mobility. Unlike fluency, which emphasizes quantity, flexibility focuses on the diversity of approaches or conceptual frameworks utilized. Scoring involves categorizing responses into distinct conceptual domains and counting the number of different categories represented. This component reflects cognitive versatility and the ability to transcend functional fixedness by approaching problems from multiple perspectives.
	
	Elaboration represents the depth and complexity of creative expression, measured through the degree of detail, development, and embellishment present in responses. In verbal creativity tasks, elaboration scores reflect the extent to which basic ideas are enhanced through additional descriptors, qualifications, or contextual specifications. For figural creativity assessments, elaboration is quantified by counting supplementary visual elements such as details, shading, color applications, and background features that extend beyond minimal task requirements. This component captures the ability to develop and refine initial creative insights into more fully realized concepts.
	
	While different creativity assessment instruments may emphasize certain components differentially, the four-factor model remains fundamental to understanding and measuring creative capabilities in both research and applied settings. The interrelationship among these components continues to be an area of active investigation, with evidence suggesting both independent and interactive contributions to overall creative performance.
	
	Automated content analysis employed a two-stage LLM-based pipeline for scalable idea extraction and semantic organization. Stage 1 utilized GPT-4.1 for structured idea extraction from user transcripts, producing JSON-formatted outputs with standardized fields (ID, title, description, evidence quotes). Stage 2 employed category induction to organize extracted ideas into coherent thematic groups ($\leq$8 categories per participant).
	
	Idea fluency showed no significant between-condition differences, though control participants generated slightly more ideas on average (Treatment M=8.55, SD=3.85; Control M=9.23, SD=3.37).
	
	Originality assessment leveraged OpenAI text-embedding-3-large representations to compute semantic distances in high-dimensional idea space. Each participant's idea portfolio was represented by the mean-pooled centroid of their normalized idea embeddings.
	
\begin{figure}
		\centering
		\includegraphics[width=0.6\linewidth]{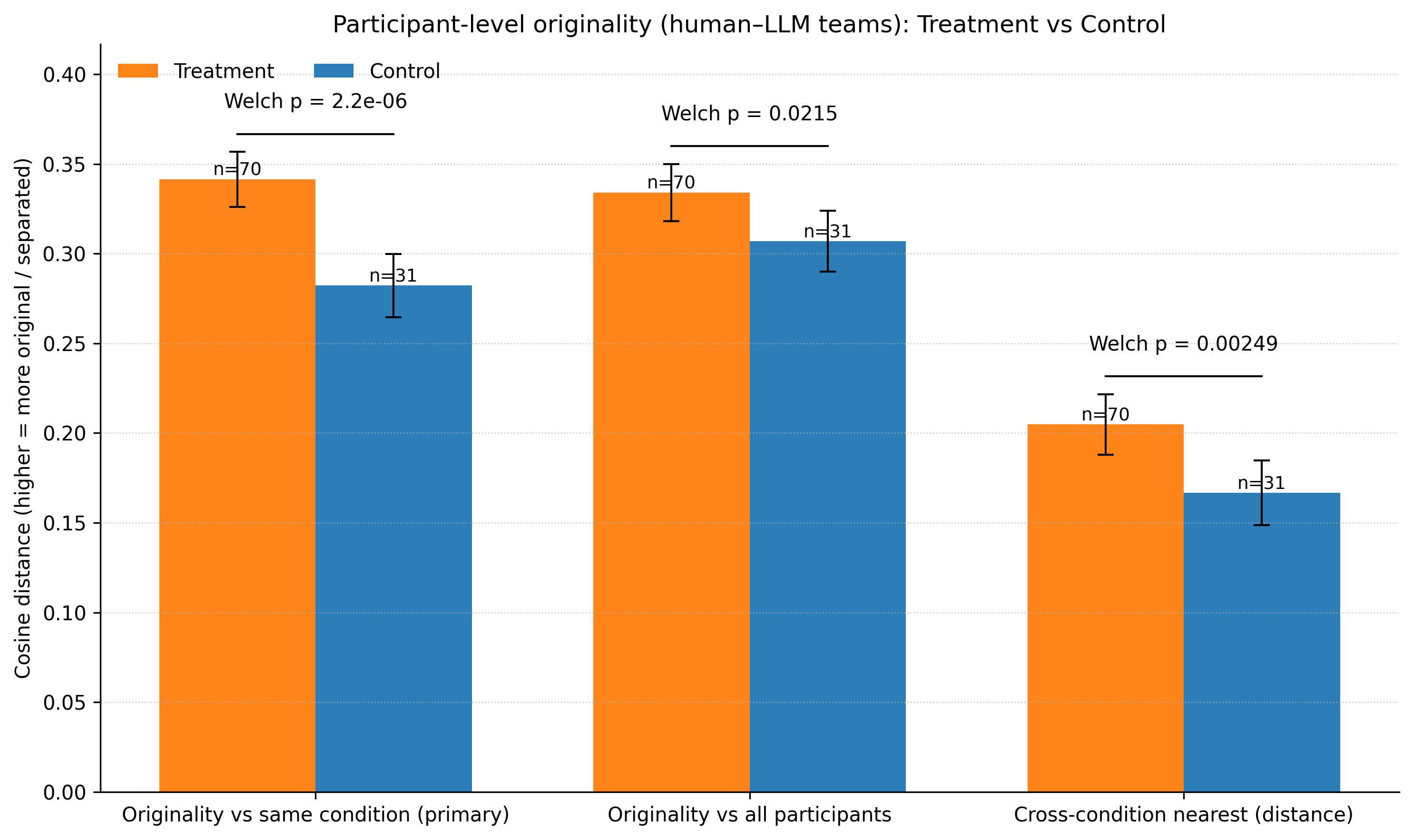}
		\caption{Participant-level originality measures (Treatment=orange, Control=blue).}
		\label{fig:originality}
	\end{figure}

	Three complementary originality measures were computed: (1) mean cosine distance to same-condition peers, (2) mean distance to all participants, and (3) nearest cross-condition neighbor distance. Treatment participants demonstrated significantly higher originality across all measures (Figure \ref{fig:originality}).
	
	Same-condition originality showed the largest effect (Treatment M=0.34, Control M=0.28; Welch's t, p=$2.20 \times\ 10^-6$), indicating that persona-guided teams occupied more distinct semantic positions. All-participant originality remained significant (Treatment M=0.33, Control M=0.31; p=.02), and cross-condition nearest neighbor distances confirmed greater between-condition separation (Treatment M=0.20, Control M=0.17; p<.01).
	
	Within-participant idea diversity showed no group differences (mean pairwise distance: Treatment M=0.59, Control M=0.59, p=.90), suggesting that originality gains resulted from between-participant differentiation rather than increased internal portfolio variation.
	
	Exploratory analysis revealed systematic relationships between idea volume, originality, and persona preference. Participants generating more ideas showed lower relative originality ($\rho$=-0.39, p<.001) but higher within-participant diversity ($\rho$=0.46, p<.001), suggesting a quantity-distinctiveness tradeoff.
	
	Within the treatment condition, convergent persona preference (Alex) correlated with greater solution ownership ($\rho$=0.25, p=.05) and fewer generated ideas ($\rho$=-0.28, p=.02). A directional but non-significant association emerged between Alex preference and participant-level originality ($\rho\approx$0.22, p$\approx$.08), consistent with convergent processing consolidating exploration into more distinctive outcomes.
	
	Together, these results demonstrate that persona-guided LLM interfaces can significantly shape creative collaboration outcomes by modulating user curiosity, perceived authorship, and semantic originality. Divergent-convergent persona differentiation promoted distinct interaction patterns and amplified creative distinctiveness at the system level without increasing within-user variance. Personality-linked engagement—particularly among agreeable, open, and conscientious users—highlights the potential of adaptive persona systems to personalize support strategies in human–AI co-creation. From an information systems perspective, these findings underscore the value of embedding psychological and behavioral modeling within generative system design to enhance both creativity and user alignment in socio-technical contexts.

	\hypertarget{discussion}{
		\chapter{Discussion}\label{discussion}}
	
	This study set out to test whether making divergent and convergent thinking explicit through two coordinated LLM personas changes both the creative process and its products. Triangulating three analytic lenses, subjective self-reports, interaction process traces, and computational semantic evaluations, provides robust evidence that persona-guided interfaces outperform conventional LLM chats in fostering creativity. Subjective questionnaires captured whom participants credited for creative help and solution authorship. Process-level data revealed staged curiosity dynamics and iterative shifts between personas that align with design thinking's iterative divergence-convergence cadence. Computational semantic analysis demonstrated that persona-guided users produced solutions occupying more original positions in semantic space without inflating idea quantity.
	
	The empirical findings reveal distinct patterns in the relationship between cognitive styles, personality traits, and user preferences for interactive systems. Divergent thinking processes demonstrate a significant positive influence on participants' subjective evaluations of solution originality and creativity, suggesting that exposure to expansive ideational approaches enhances the perceived innovative quality of generated outcomes. This phenomenon aligns with theoretical frameworks positing that divergent cognitive strategies facilitate recognition and appreciation of novel conceptual combinations.
	
	Furthermore, individual differences in personality dimensions emerge as critical determinants of interaction preferences. Participants scoring high on the openness to experience dimension exhibit pronounced preferences for innovative interaction modalities relative to conventional chatbot interfaces, a finding consistent with their dispositional tendency toward novelty-seeking and experimental engagement. This preference pattern reflects the fundamental alignment between trait openness and receptivity to unconventional technological interfaces.
	
	Conversely, conscientious individuals demonstrate systematic preferences for convergent thinking personas, a pattern that corresponds with their characteristic orientations toward systematic organization, methodical processing, and comprehensive task completion. The attraction to convergent approaches among highly conscientious users likely stems from the congruence between structured problem-solving methodologies and their inherent predisposition toward orderliness and thorough preparation.
	
	Similarly, neuroticism emerges as a significant predictor of convergent solution preferences, with individuals exhibiting elevated neuroticism scores displaying marked tendencies to favor interactions culminating in definitive, closed-form solutions. This preference pattern may reflect the anxiety-reducing properties of conclusive outcomes for individuals with heightened emotional reactivity, suggesting that convergent approaches provide psychological closure that aligns with their need for certainty and resolution in problem-solving contexts.

	Participants more frequently perceived the divergent persona, Taylor, as the primary creative contributor despite controlled blinding against visual biases such as neutral names and randomized button positions. This attribution likely arises from Taylor's salient expansions and unique idea combinations that are easier to detect and remember compared to Alex's convergent, evaluative role. Alex's narrowing contributions arguably function as a foil that accentuates the originality of Taylor’s proposals through contrast. These findings strengthen the inference that perceived creativity differences reflect genuine cognitive role dynamics rather than superficial interface effects.
	
	Importantly, the creative process was not a rigid sequence strictly separating divergence and convergence. Instead, participants fluidly interleaved interactions with both personas, consistent with Goldschmidt's notion of linkography describing rapid forelinking and backlinking in creative cognition. The interface thus provides visible, user-initiated control over mode shifts without constraining the iterative, intertwined texture of real-world creative problem-solving.
	
	While overall creativity ratings were comparable across conditions, within-person difference scores revealed that participants consistently credited Taylor more for creativity compared to Alex within the same session. This pattern suggests that convergent persona guidance serves as an evaluative anchor calibrating novel idea thresholds without inflating global creativity perceptions. From a design perspective, this indicates that well-timed convergent input allows divergent contributions to stand out, improving user recognition of novelty.
	
	These insights directly inform design principles for AI creativity support systems. Persona design must involve concrete, visible expansions from the divergent partner and focused decision-making prompts from the convergent partner. Interface designs preserving simultaneous visibility and easy role switching support natural inquiry choreography, where exploratory questions peak early with Taylor and focused structuring with Alex rises mid-session. Moreover, light-touch personalization leveraging trait-linked engagement patterns can guide users to the appropriate cognitive stance—widening or narrowing—while maintaining user autonomy. This approach resonates with recent HCI findings that effective persona and role-prompting configurations enhance creative behavior without hindering user independence \citep{kumar2025human, muller2024group, shaer2024ai}.
	
	Finally, this work contributes to the debate on AI coaching versus direct answer giving. While prior large-scale studies highlight risks of detrimental dependence from coaching, our findings nuance this view: a single-session design integrating dynamic, role-specific personas can balance focus and creativity, preserving originality while encouraging ownership and reducing reliance. This highlights the importance of calibrating “coach” behaviors and user control mechanisms in socio-technical AI design.
	
	\section{Limitations}
	Several limitations deserve acknowledgement. The study employed the BFI-2-XS, a brief Big Five personality measure trading precision for brevity, likely reducing sensitivity to nuanced trait-persona interactions. The information-seeking behavior metric relied on counting question marks as a proxy for questions; some questions may not use punctuation while some marked questions may not represent true inquiries, warranting more refined definitions in future analyses.
	
	The experimental design focused on a single creative problem prompt and one LLM family during one session, limiting generalizability across problem types, longer collaborations, or cross-model robustness. The objective originality measure depended on embedding-based semantic distances combined with an LLM-aided idea extraction pipeline,an approach that, while scalable and broadly aligned with qualitative differences, requires triangulation with blinded human evaluations and alternative embeddings for convergent validity.
	
	\section{Future Work}
	Building on this foundation, future research should enrich measurement by augmenting idea annotation pipelines with multiple agent flows, comparing different LLM families, and pairing embedding-based metrics with expert creativity ratings to better validate semantic originality assessments.
	
	Interaction design studies should explore factorial variations in persona availability, role orchestration, and longer collaborative sequences alternating coached and autonomous rounds to examine effects on creativity persistence and learning transfer. Incorporating qualitative user feedback and think-aloud methods could deepen understanding of user experience and cognitive strategies during persona-guided collaboration.
	
	Addressing language diversity by extending persona designs for multilingual contexts promises broader applicability. Finally, further investigation into trust calibration, authorship attributions, and ethical governance will be vital to responsibly deploying persona-guided creativity support at scale.
	
	\hypertarget{conclusion}{
		\chapter{Conclusion}\label{conclusion}}
	This paper evaluated a persona-guided LLM chat interface designed to render divergent and convergent thinking explicit, separable, and user-orchestrated across a creative problem-solving task. By externalizing roles and enabling smooth switches, the interface facilitated a natural progression from exploration to framing to evaluation, mirroring design thinking’s iterative cycles.
	
	Users staged their curiosity accordingly: exploratory questions increased with the divergent persona early on, shifting toward structured, prioritizing queries with the convergent persona as sessions progressed. Participants perceived the divergent persona as the primary source of novelty, while reliance on the convergent persona aligned with higher ownership and fewer, more focused ideas. These process signatures corresponded with computational evidence of increased originality without inflating idea quantity, demonstrating that clear role delineation and well-timed handoffs can effectively scaffold search and solution consolidation while keeping control firmly with the user.
	
	Overall, this work suggests a practical interaction design paradigm: treat LLMs as complementary creative partners with clear, explicitly communicated cognitive roles and afford users timely, lightweight mode switches. Supporting brief, user-owned commitments during convergence preserves authorship and enhances collaboration. While constrained by a single task domain and short personality inventory, converging behavioral, experiential, and computational signals establish a concrete design lesson for human-AI creativity support: explicit role clarity, phased cognitive mode shifts, and light personalization collectively empower teams to explore broadly, converge deliberately, and maintain authorship.

	\pagebreak
	\printbibliography
	
\end{document}